\documentclass[a4paper,12pt, english,german]{article}
\usepackage[a4paper, total={6in, 9.5in}]{geometry}
\usepackage{mathtools}
\usepackage{graphicx}

\usepackage{amssymb,latexsym,amsmath,color,mathrsfs,ifsym,graphics,stmaryrd} 
\usepackage[colorlinks,linkcolor=blue,urlcolor=blue,citecolor=black,
plainpages=false,pdfpagelabels,breaklinks]{hyperref}

\title{The  Relativistic Transactional Interpretation: Immune to the Maudlin Challenge}

\begin{document}  
\maketitle

\centerline{Ruth E. Kastner}

\bigskip
\begin{abstract}
\noindent
The Transactional Interpretation has been subject at various times to a challenge based on a type of thought experiment first proposed by Maudlin. It has been argued by several authors that such experiments do not in fact constitute a significant problem for the transactional picture. The purpose of this work is to point out that when the relativistic level of the interpretation is considered, Maudlin-type challenges cannot even be mounted, since the putative `slow-moving offer wave,' taken as subject to contingent confirmation, does not exist. This is a consequence of the Davies relativistic quantum-mechanical version of the direct-action theory together with the asymmetry between fermionic field sources and bosonic fields. The Maudlin challenge therefore evaporates completely when the relativistic level of the theory is taken into account.

\end{abstract}

\bigskip

\section{The Basics: A brief review}

The Transactional Interpretation (TI), first proposed by John Cramer \cite{Cramer}, is based on the direct-action theory of electromagnetism by Wheeler and Feynman \cite{WF}. A relativistic extension of TI has been developed by the present author; that is based on Davies' direct-action theory of quantum electrodynamics \cite{Davies}. (See also \cite{KastnerEJTP}.)  Due to its possibilist ontology, that model has been termed `PTI' \cite{Kastner2012}, but the important feature is its relativistic nature, which provides further clarification of the conditions for emission and absorption. Therefore, for purposes of this discussion and going forward, I will refer to that model as the Relativistic Transactional Interpretation, RTI.

First, some terminology: in TI and RTI, the usual quantum state $| \Psi \rangle$ is called an `offer wave' (OW), and the advanced response $\langle a|$ of an absorber $A$ is called a `confirmation wave' (CW). In general, many absorbers $A,B,C,....$ respond to an OW, where each absorber responds to the component of the OW that reaches it. The OW component reaching an absorber $X$ would be $\langle x | \Psi \rangle |x\rangle$, and it would respond with the adjoint (advanced) form $\langle x | \langle \Psi |x\rangle$. The product of these two amplitudes corresponds to the final amplitude of the `echo' of the CW from $X$ at the locus of the emitter (this was shown in \cite{Cramer}) and reflects the Born Rule as a probabilistic weight of the `circuit' from the emitter to absorber and back, the latter being called an \textit{incipient transaction}. Meanwhile, the sum of the weighted outer products (projection operators) based on all CW responses--each representing an incipient transaction--constitutes the mixed state identified by von Neumann as resulting from the non-unitary process of measurement (\cite{Kastner2012}, Chapter 3).  Thus, TI provides a physical explanation for both the Born Rule and the measurement transition from a pure to a mixed state. The additional step from the mixed state to the `collapse' to just one outcome is understood in RTI as an analog of spontaneous symmetry breaking; the `winning' transaction, corresponding to the outcome of the measurement, is termed an \textit{actualized transaction}. The absorber that actually receives the quantum is called the \textit{receiving absorber}. This is to emphasize that other absorbers participate in the process but do not end up receiving the actualized quantum.
 
The other feature of this process, which gives it its possibilist ontology, is that the quantum entities (OW, CW, virtual quanta) are all pre-spacetime objects: Heisenbergian \textit{potentiae} (see, e.g., \cite{KKE}). Spacetime events only occur as a final result of OW/CW negotiations, resulting in collapse to an actualized transaction. Thus, the collapse is not something that happens \textit{within} spacetime; rather, collapse is the process of spacetime emergence. Specifically, what emerges as a result of collapse is the  emission event, the absorption event, and their connection via the exchanged quantum (see \cite{Kastner_causal}). It is only upon actualization of the transaction that a real quantum is emitted and absorbed at the receiving absorber.\footnote{Maudlin is thus quite correct when he says: ``It is also notable that in the electromagnetic case the relevant fields are defined on, and propagate over, space-time. The wave-function is defined on configuration space. Cramer does not seem to take account of this, writing always as if his offer and confirmation waves were simply being sent through space. Any theory which seeks to make the wave-function directly a medium of backwards causation ought to take this into account.''\cite{Maudlin}, p. 203. This issue is corrected in RTI, although the `backwards causation' is understood in a new way, as part of the establishment of a spacetime interval rather than an influence propagating within a background spacetime.}

Now let us briefly review the Maudlin thought experiment (\cite{Maudlin}, p. 200; see Figure 1). It envisions a `slow-moving OW' (assumed traveling at speed $v<c$) emitted at $t=0$ in a superposition of rightward and leftward momentum states. On the right at some distance $d$ is a fixed detector R, and positioned behind R (initially on the right) is a moveable detector L. If, after a suitable time has passed ($t_1=\frac{d}{v}$), there is no detection at R, L is quickly swung around to intercept the OW on the left, where (so the proposal goes) a left-hand CW is generated and the particle must be detected at L with certainty. Thus, this is intended to be a `contingent absorber experiment' (\cite{Kastner2012}, Chapter 5]): it is assumed that the existence of a confirmation from the left-hand side is contingent on the transaction between the source and R failing.

\begin{figure}[h]
\centering
\includegraphics[width=3.0in]{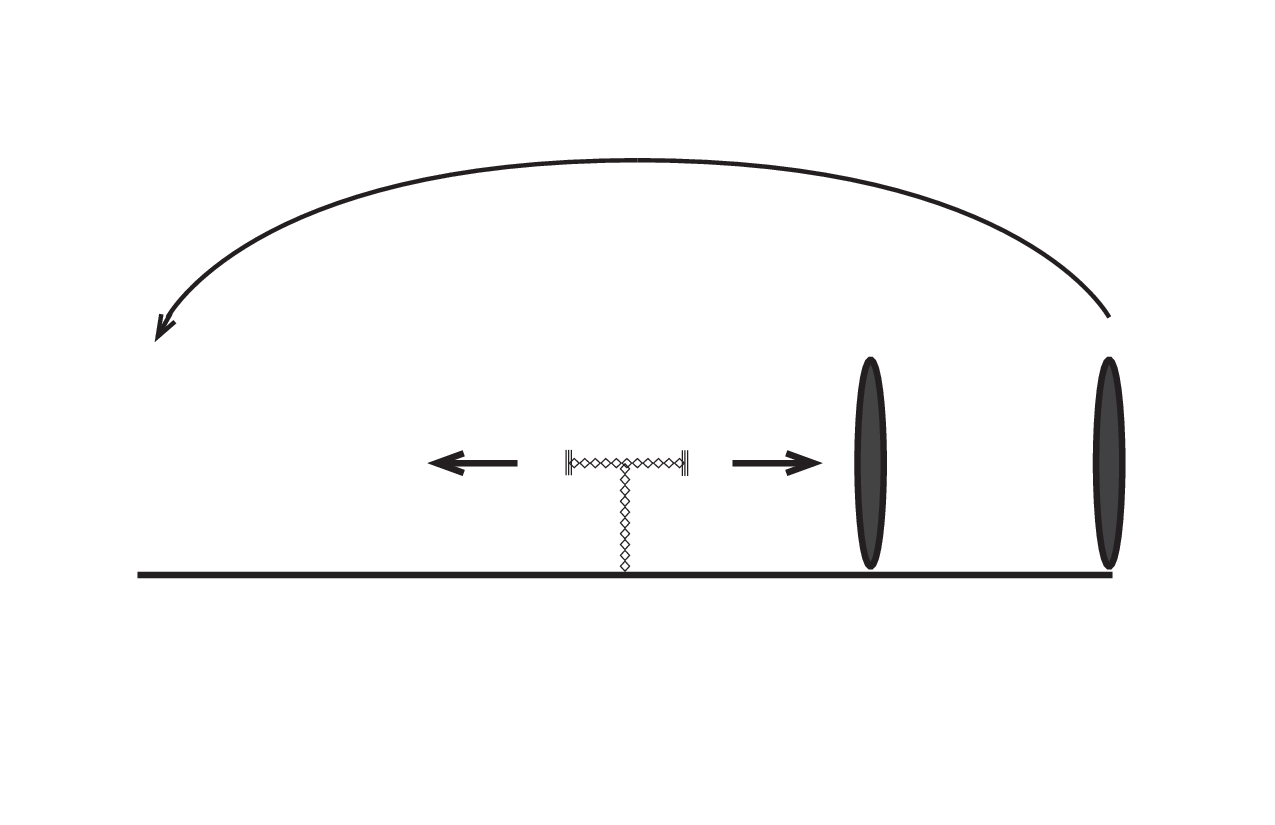}
\begin{center}
\caption{\small{Maudlin Thought Experiment}}
\label{default}
\end{center}
\end{figure}

\newpage
Maudlin's intent was to provide a counterexample to the picture provided in \cite{Cramer}, in which there are well-defined OW/CW matchups for all possible detection sites. The challenge presented for the original TI was twofold: (i) the probability of 1/2 for the leftward transaction was thought to be inconsistent with the fact that whenever the left-hand CW was present that transaction would always be actualized; and (ii) the situation at $t=0$ appeared ill-defined, since (if the CW is really contingent as imagined) it is uncertain whether or not the (backward-evolving) CW will be emitted from the left.

Both these concerns have been addressed and resolved elsewhere (cf. \cite{Marchildon}, \cite{Kastner2012}, Chapter 5). However, these responses assumed that the Maudlin experiment could in-principle be carried out. The purpose of this paper is to observe that in fact this is not the case; no such experiment can actually be done, and therefore the challenge disappears.

\section{Applicability of the `offer wave' concept}

 \indent   It has been shown that any quantized field theory can formally be re-expressed as a direct action theory \cite{Narlikar}. However, owing to the specifics of the interacting fields in our world, RTI has intrinsic restrictions on what kinds of entities constitute offer waves. That is, some objects describable as quantum systems, such as atoms, do not constitute offer waves, in that they are not excitations of a specific quantum field--instead, they are bound states \cite{Kastner_bound}. Quanta of fermionic fields (as opposed to photons) could be considered a type of `offer wave'; but as they are really emitters and absorbers (e.g., electrons), they participate in actualized transactions indirectly, through confirmations of the products of their interactions, rather than by generating confirmations themselves. This work discusses these situations, and then applies the findings to the Maudlin challenge to see why it cannot be mounted. 
 
 First, as indicated above, the `offer wave' concept refers only to the excited states of a quantum field.  A specific example would be a one-photon Fock state $| k \rangle$. On the other hand, if the system at hand is not a specific field excitation of this sort, even though it may still be described by an effective quantum state, it is not an offer wave. It therefore does not generate a corresponding confirmation wave. As noted above, an example of such a system would be an atom, which is a bound state of several different quantum fields as opposed to an excitation of a single quantum field.
 
At this point the relevance for the Maudlin challenge is already evident: the latter proposes a `slow-moving quantum' subject to contingent absorption. The `slow-moving quantum' cannot be anything other than a field excitation for a quantum with nonvanishing mass if it is to be eligible for `offer wave' status, so an atom cannot instantiate the experiment.  In order to obtain an offer corresponding to a subluminal quantum, one must use a matter field, such as the Dirac field. But the latter will be a source of bosonic fields--i.e., an emitter/absorber, which brings us to the second important point: the asymmetry between field sources and their generated fields gives rise to a situation in which a field source participates in transactions indirectly, by way of its emitted field. Thus, it turns out that even a fermionic field excitation does not really count as an `offer wave', if by that, we mean something that requires its own matching confirmation wave in order to be detected. In what follows, we consider this issue in further detail. 
 
\section{Field sources are deteced without matching confirmations}

 In quantum electrodynamics, the (fermionic) Dirac field is the source of the (bosonic) electromagnetic field, but the following considerations apply to any quantum field and its sources. It is well known that in  interactions between fields, the field source has a different physical character from the field of which it is a source. This distinction is reflected in the fact that gauge bosons are the force carriers, as opposed to the fermionic matter fields which are sources of gauge bosons. The asymmetry in question is exhibited for example in the basic QED vertex, which has only one photon line, plus an incoming and outgoing fermion line, due to the nature of the coupling between the Dirac Field and the electromagnetic field, given by $eA_\mu \times \bar{\Psi} \gamma^{\mu} \Psi$. 

\begin{figure}[h]
\centering
\includegraphics[width=2.5in]{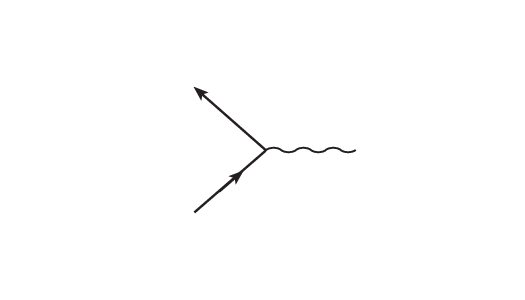}
\begin{center}
\caption{\small{QED Vertex}}
\label{default}
\end{center}
\end{figure}

 \begin{figure}[h]
 \centering
\includegraphics[width=4in]{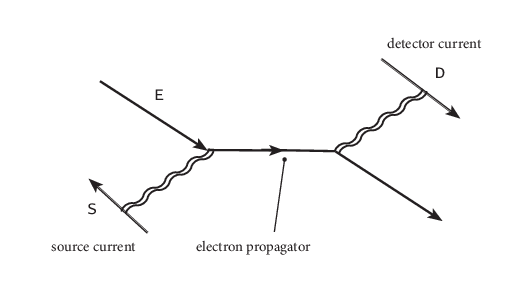}
\begin{center}
\caption{\small{Electron Detection}}
\label{default}
\end{center}
\end{figure}

Due to this asymmetry, not all `offer waves' generate their own confirmations when participating in transactions. Fermionic field sources participate in transactions indirectly, by way of confirmations of the fields of which they are a source.\footnote{Even if fermionic quantized fields can be formally recast as direct-action fields, only bosonic fields (subject to a `gauge field' description) engage in transactions by way of their own confirmations. The deep physical meaning behind this, relevant for the RTI picture of spacetime emergence from actualized transactions, is that only the bosonic fields correspond to spacetime symmetries.} For example (see Figure 2), an electron is liberated from a bound state by absorbing an incoming photon from another charged current S, and then emits a photon offer wave, which is confirmed by another charged source field D (typically an electron) in the detector.\footnote{We recall here that at the quantum relativistic level, photon OW and CW are mutually generated; this is the equivalent of the 'light tight box' condition, but it requires no special cosmological boundary condition. The mutuality is established through virtual photon activity, and there is never an OW without corresponding CW.} The outgoing electron  is actualized, as an emitter, in a particular state as well, even though it was not confirmed by any `electron CW.' (This is discussed further below.) The Born probabilities describing the electron's possible outgoing states arise from its participation in the associated photon transactions which enable its (indirect) detection. (See \cite{KastnerCramer} for details.)

The asymmetry between the fermionic field source (the electron E) and its emitted/absorbed fields (the photon lines) is evident here in that it is an electron propagator that connects the two interaction vertices.\footnote{For real processes of ionization (liberation of an electron for use in an experiment) and radiative recombination, we have a real, on-shell electron, corresponding to the pole in the Feynman propagator; see \cite{Davies}, 1972.} This allows the electron E to be indirectly actualized via its interaction with the electromagnetic field.  Thus, upon detection of the emitted photon by D, the electron E is actualized without having generated its own confirmation. Instead, it has acted as an emitter in a photon transaction, and has been actualized as the site of an emission event. The doubled photon lines indicate that a photon CW is generated at each end of the process: the electron acts as an absorber in its initial liberation from a bound state, and then as an emitter in its final detection via radiative recombination.\footnote{In between, the liberated electron can be subject to unitary interactions which place it in a superposition of leftward and rightward momentum states. But since neither of these require a matching `electron confirmation' for their detection, there is no problem.}

\section{Why the Maudlin challenge evaporates}

Returning now specifically to the Maudlin experiment: in order for the `slow-moving quantum' to be considered an `offer,' it would have to be a non-composite matter field excitation of some sort, such as an electron state $|\psi>$. But the latter is a source of a bosonic field (the electromagnetic field), which is the mediator of electron detection, as described above.\footnote{The weak field is a massive boson, but its range is far too short to be useful for the Maudlin experiment.} Recall that a free electron can neither emit nor absorb a real photon offer wave (due to energy conservation). An electron subject to detection is always liberated from some bound state (by absorbing electromagnetic energy) and detected via its becoming part of a new bound state (by emitting electromagnetic energy), not through being confirmed by a matching `electron confirmation.' 

 To see this, refer again to Figure 2.  On the incoming side on the left, a photon offer is emitted from another charged current (labelled S) and confirmed/absorbed by the electron E, which is thereby liberated from its initial bound state. (Assuming we indeed have energy conservation, i.e., that these are not virtual processes, a typical example of such a process is the photo-electric effect.) Meanwhile, at the outgoing side, E  emits a photon offer. The latter is confirmed by a charged current D in the detector, and the electron becomes bound in some energy level in the detector. (This is a process of `radiative recombination'.) The actualization of the emissions and absorptions of each photon actualize E in a particular state as an emitter/absorber, since (recall from Section 1) all actualized transactions actualize three things: the emission event, the absorption event, and the transferred photon.  As noted above, the outgoing electron (actualized in the particular emission state that gave rise to the outgoing photon detection) becomes incorporated into another bound state (such as a conduction band in a metal), rather than prompting its own confirmation at the Dirac field level. Thus, the Maudlin challenge lacks a `slow-moving offer wave' that requires a matching confirmation for detection, and cannot be mounted at all.

\section{Conclusion}

Recent developments of the Relativistic Transactional Interpretation (RTI) have been presented which nullify the Maudlin challenge for the Transactional Interpretation (TI). These new development are: (1) offer waves are excitations of quantum fields, so slow-moving composite quantum objects such as atoms are not eligible for the experiment; (2) fermionic matter fields describable as field excitations are not detected by way of their own matching confirmations, but by transactions involving their emitted/absorbed bosonic fields. These developments result in the evaporation of the Maudlin challenge, since there is no `slow-moving offer wave' to begin with, unless it is a non-composite fermion such as an electron, but that does not require a matching confirmation to be detected, so the term 'offer wave' is probably not even appropriate for fermionic fields. In this case, an electron `offer' is detected by way of photon transactions, of which it is a source. Thus, there would therefore never be a  'slow-moving' contingent CW situation--i.e., never a situation in which a required CW would only generated based on some prior non-detection. Finally, these observations should not be mistaken as \textit{ad hoc} maneuvers to evade the Maudlin challenge; rather, they arise directly from the Davies QED absorber theory upon which RTI is based, but had not been previously taken into account. 

\bigskip

Acknowledgments. The author is grateful to an anonymous referee for helpful suggestions for improvement of the presentation.


\newpage

\begin{thebibliography}{10}



\bibitem{Cramer}
Cramer J G.
\newblock The Transactional Interpretation of Quantum Mechanics.
\newblock \emph{Reviews of Modern Physics 58}, 647-688, 1986.

\bibitem{WF} Feynman, R P and Wheeler, J A.
\newblock ``Interaction with the Absorber as the Mechanism of Radiation",
\newblock\emph{Reviews of Modern Physics, 17} 157-161 (1945); 
\newblock and 	``Classical Electrodynamics in Terms of Direct Interparticle Action",
\newblock \emph{Reviews of Modern Physics 21}, 425-433 (1949).


\bibitem{Davies} Davies, PCW.
\newblock ``Extension of Wheeler-Feynman Quantum Theory to the Relativistic Domain I. Scattering Processes", 
\newblock \emph{ J. Phys. A: Gen. Phys. 4}, 836 (1971); 
\newblock and ``Extension of Wheeler-Feynman Quantum Theory to the Relativistic Domain II. Emission Processes",
\newblock \emph{J. Phys. A: Gen. Phys. 5}, 1025-1036 (1972).

\bibitem{KastnerEJTP}
Kastner, R.E.
\newblock ``On Real and Virtual Photons in the Davies Theory of Time-Symmetric Quantum Electrodynamics."
\newblock \emph{Electronic Journal of Theoretical Physics 11}, 30: 75-86 (2014).
\newblock http://www.ejtp.com/articles/ejtpv11i30p75.pdf

\bibitem{Kastner2012}
Kastner R E.
\newblock \emph{The Transactional Interpretation of Quantum Mechanics: The Reality of Possibility}.
\newblock Cambridge: Cambridge University Press (2012).

\bibitem{KKE}
Kastner, R.E., Kauffman, S., Epperson, M.
\newblock ``Taking Heisenberg's Potentia Seriously.''
\newblock\emph{International Journal of Quantum Foundations 4},2; 158?172 (2018).

\bibitem{Kastner_causal}
Kastner R E.
\newblock ``The Emergence of Spacetime: Transactions and Causal Sets," in Licata, I. (Ed.),
\newblock \emph{Beyond Peaceful Coexistence: The Emergence of Space, Time and Quantum}. London: Imperial College Press (2016).
\newblock Preprint version: https://arxiv.org/abs/1411.2072

\bibitem{Maudlin}
Maudlin T.
\newblock \emph{Quantum Nonlocality and Relativity}, 3rd Edition.
\newblock Oxford: Blackwell (2011). pp. 184-185.

\bibitem{Marchildon} 
Marchildon, L.
\newblock ``Causal Loops and Collapse in the Transactional Interpretation of Quantum Mechanics" 
\newblock \emph{Physics Essays 19}, 422-9 (2006)

\bibitem{Narlikar}
Narlikar, J. V. 
\newblock ``On the general correspondence between field theories and the theories of direct particle interaction.''
\newblock \emph{ Proc. Cam. Phil. Soc. 64}, 1071 (1968).

\bibitem{Kastner_bound}
Kastner, R.E.
\newblock ``Bound States as Emergent Quantum Structures." in Kastner, R.E., Dugic J., and Jaroszkiewicz, G (Eds.),
\newblock \emph{Quantum Structural Studies}. Singapore: World Scientific (2016).
\newblock Preprint version: https://arxiv.org/abs/1601.07169

\bibitem{KastnerCramer}
Kastner, R.E. and Cramer, J.G.
\newblock ``Quantifying Absorption in the Transactional Interpretation."
\newblock\emph{International Journal of Quantum Foundations 4},2: 210-222 (2018). 



\end{thebibliography}
\end{document}